\documentclass[a4paper,11pt]{article}
\begin{document}
\def\boxit#1{\vcenter{\hrule\hbox{\vrule\kern8pt
      \vbox{\kern8pt#1\kern8pt}\kern8pt\vrule}\hrule}}
\def\Boxed#1{\boxit{\hbox{$\displaystyle{#1}$}}} 
\def\sqr#1#2{{\vcenter{\vbox{\hrule height.#2pt
        \hbox{\vrule width.#2pt height#1pt \kern#1pt
          \vrule width.#2pt}
        \hrule height.#2pt}}}}
\def\square{\mathchoice\sqr34\sqr34\sqr{2.1}3\sqr{1.5}3}
\def\Square{\mathchoice\sqr67\sqr67\sqr{5.1}3\sqr{1.5}3}
\def\AJP{{\it Am. J. Phys.}}
\def\AM{{\it Ann. Math.}}
\def\AP{{\it Ann. Phys.}}
\def\CQG{{\it Class. Quantum Grav.}}
\def\GRG{{\it Gen. Rel. Grav.}}
\def\JMP{{\it J. Math. Phys.}}
\def\JP{{\it J. Phys.}}
\def\JSIRAN{{\it J. Sci. I. R. Iran}}
\def\NC{{\it Nuovo Cim.}}
\def\NP{{\it Nucl. Phys.}}
\def\PL{{\it Phys. Lett.}}
\def\PR{{\it Phys. Rev.}}
\def\PRL{{\it Phys. Rev. Lett.}}
\def\PRp{{\it Phys. Rep.}}
\def\RMP{{\it Rev. Mod. Phys.}}
\title{About Gravitomagnetism}
\author{{\small Behrooz Malekolkalami}\footnote{e--mail: b\_malekolkalami@sbu.ac.ir}\ \ {\small and
         Mehrdad Farhoudi}\footnote{e--mail: m-farhoudi@sbu.ac.ir}\\
        {\small Department of Physics, Shahid Beheshti University, G. C.,}\\
        {\small Evin, Tehran 19839, Iran}}
\date{\small September 2, 2008}
\maketitle
\begin{abstract}
The gravitomagnetic field is the force exerted by a
\textit{moving} body on the basis of the intriguing interplay
between geometry and dynamics which is the analog to the magnetic
field of a moving charged body in electromagnetism. The existence
of such a field has been demonstrated based on special relativity
approach and also by special relativity plus the gravitational
time dilation for two different cases, a moving infinite line and
a uniformly moving point mass, respectively. We treat these two
approaches when the applied cases are switched while appropriate
key points are employed. Thus, we demonstrate that the strength of
the resulted gravitomagnetic field in the latter approach is twice
the former. Then, we also discuss the full linearized general
relativity and show that it should give the same strength for
gravitomagnetic field as the latter approach. Hence, through an
exact analogy with the electrodynamic equations, we present an
argument in order to indicate the best definition amongst those
considered in this issue in the literature. Finally, we
investigate the gravitomagnetic effects and consequences of
different definitions on the geodesic equation including the
second order approximation terms.
\end{abstract}
\bigskip
\section{Introduction}
\indent

The analogous idea of the electric theory and the Newtonian
gravitational theory inspiring a Maxwell--type gravitational
theory is dated back to the second half of the nineteenth
century~\cite{19thcentury1}--\cite{19thcentury4}. This idea which
was also explored by Einstein~\cite{einstein13} was revived and
extended by Sciama~\cite{sciama53,sciama59}. Though, his theory
was unfruitful because, unlike the electric charge, the mass
charges are~not invariant nor additive~\cite{rindler79}. The
properties which are actually due to the linearity of
Maxwell--type theories, as, the linearized weak field
approximation of the Einstein gravitational theory presented to
second order approximation does~\cite{t18v21}--\cite{sch84}. This
issue has been demonstrated in several books as well, see, e.g.,
Refs.~\cite{sciama59,rindler79,ciuwheel95} and references therein.

Indeed, the introduction of a gravitomagnetic~({\bf GM}) field is
unavoidable when one brings the Newtonian gravitational theory and
the Lorentz invariance together in a consistent framework. This
effect is the usual Maxwellian feature which has Machian root,
see, e.g., Ref.~\cite{lichmash04}. Dynamical equations for a weak
gravitational field similar to the Maxwell equations has been
deduced~\cite{braginsky-etal77} by the parameterized
post--Newtonian formalism. Also, more attention has been
made~\cite{harris91} in the analogy between general relativity and
electromagnetism for slowly motion in a weak gravitational field.

However, there are other dissimilarities. For example, not only
negative mass charge has~not been detected yet, the \emph{like}
mass charges attract rather than repelling each others. Issues
which, partially, have been stated as the weak version of the
principle of equivalence, that is, the gravitational field couples
to everything, and or, all forms of energy acts as sources of
gravitational field.

Nevertheless, in the last two decades, extensive attention, has
been taken on this issue, see, e.g.,
Refs.~\cite{rugtar02}--\cite{mash2003} and references therein.
Especially, and contrary to assertion made in
Ref.~\cite{national86}, the existence of GM interaction has been
claimed in 1988~\cite{nordt88,mashpw89}, and evidence for the GM
field has been suggested, see
Refs.~\cite{ciuwheel95,rugtar02,lim02,wz05,cp06} and references
therein.

The introduction of GM field as an analogy to the magnetic field
comes about from the need to know the force exerted by a
\emph{moving} body on the basis of the intriguing interplay
between geometry and dynamics, as emphasized by
Sciama~\cite{sciama59}. A fundamental idea that perhaps motivated
the two, almost recent, articles,
Refs.~\cite{bedfordkrumm85,kolbenstvedt88}, to present the
existence of gravitomagnetism. Indeed, in
Ref.~\cite{bedfordkrumm85}, based on special relativity~({\bf
SR}), the existence of GM field has been shown for a moving
infinite line\footnote{In practice a very long moving line.}~({\bf
MIL}) of constant mass charge density, analogous to the magnetic
field from a straight current. Whereas, in
Ref.~\cite{kolbenstvedt88}, again based on SR plus the aid of the
gravitational time dilation, the existence of GM field has been
demonstrated for a uniformly moving point mass~({\bf MPM}) by
considering its line element and comparing the resulted Lagrangian
with the corresponding non--relativistic electromagnetic~({\bf
EM}) case.

The purpose of this article is to study the correspondence of the
two approaches employed in
Refs.~\cite{bedfordkrumm85,kolbenstvedt88} and compare them with
the results of the full linearized general relativity~({\bf LGR}).
Hence, in the next section, we briefly iterate these two
approaches in a more elegant and consistent manner while we switch
their applied cases and employ appropriate key points in each
condition. Implicitly, we refer to these two approaches as SR and
semi SR~({\bf SSR}) approaches, respectively. In Section~3, we
apply the LGR to general cases, then we compare this approach with
the previous results and give a short discussion and suggestion in
order to indicate the best definition amongst those used for this
issue in the literature. Then, in Section~4, we investigate the GM
effects and consequences of different definitions on the geodesic
equation including the second--order approximation terms. Finally,
a brief conclusion is given in the last section.

The necessary calculations have also been furnished in the
Appendix at the end of the article. We use an isotropic
space--time of signature $-2$ and set $c=1$. Also, we employ the
convention that lower case Latin indices run from zero to three,
whereas the lower case Greek indices run from one to three.

\section{Switching Line and Point Mass Cases}
\indent

In order to study the correspondence of the two approaches
employed in Refs.~\cite{bedfordkrumm85,kolbenstvedt88}, we iterate
them consistently while we switch the applied cases and use an
appropriate key point in each condition.

Firstly, following the approach of Ref.~\cite{bedfordkrumm85}, we
consider a MPM with rest mass $M_o$ and constant linear velocity
${\bf v}$ with respect to a frame of reference, say $S$, for
simplicity along the positive $x$--direction. Also, suppose a
point test mass\footnote{A test mass is obviously a mass which
experiences a gravitational field but does not itself alter the
field or contribute to the field.}\
 with rest mass $m_o$ moving under the
influence of $M_o$, which, without loss of generality, we assume
has the instantaneous 3--velocity ${\bf u}'=(0, u'_y, 0)$, where
$u'_y\ll 1$, in the rest frame of $M_o$, $S'$, in $x'y'$ plane.
That is, we consider the orthogonal relative motion of two bodies.

In this rest frame, the 3--force on $m_o$ is ${\bf f}'=(0, f'_y,
0)$, where $|f'_y|=GM_om/r'^2$ and $r'$ is the instantaneous
proper distance from $m_o$ to $M_o$. Using the standard Lorentz
transformation for 4--force, $F^a=\gamma_u \bigl({\bf u}\cdot {\bf
f}\, ; {\bf f}\bigr)$, and $m\simeq m_o$ to the first--order
approximation, one gets the 3--force on $m_o$ in the frame $S$ to
be
\begin{equation}\label{F1}
{\bf f}=\Bigl(vu'_yf'_y, {f'_y\over \gamma},
0\Bigr)=\mp\Bigl(v\gamma u_y\frac{GM_om_o}{r^2},
\frac{GM_om_o}{\gamma r^2}, 0\Bigr)\,,
\end{equation}
where obviously $\gamma\equiv\gamma_v=1/\sqrt{1-v^2}$, $r'=r$, for
a perpendicular direction of motion, and the negative/positive
sign is for when $m_o$ is above/under $M_o$, which is assumed to
be on the $x'$--axis. Also, note that, obviously $u_x=v$ in this
situation.

Now, as mentioned in the introduction, magnetic type forces can be
interpreted as relativistic forces. That is, if all one knows in
the EM is the Coulomb law, then, by using the SR and the
invariance of charge, one can demonstrate that a magnetic field,
which exerts the Lorentz force on a test charge, must
exist~\cite{res68}. Indeed, the force $f_x$, which is needed to
keep the velocity of $m_o$ in this direction as a constant value,
is actually the gravitational analog of the magnetic force, that
is $f^{(\rm gm)}_x$. In Ref.~\cite{bedfordkrumm85}, where the MIL
case has been applied, the length contraction has been employed in
order to proceed. In our situation, an appropriate key point is
the concentration of the gravitational field lines in the
transverse direction, see, e.g., Ref.~\cite{res68}. In order to
find the gravitational analog of the electric force on $m_o$ in
$S$, we use the relation derived in various textbooks, e.g.
Ref.~\cite{res68}, for the electric field of a moving charge,
therefore\footnote{The gravitoelectric~({\bf GE}) notation ${\bf
E}^{(\rm ge)}$, as the gravity analog of the electric field, has
been used instead of the usual ${\bf g}$ field.}
\begin{equation}\label{F2}
{\bf f}^{(\rm ge)}=m_o{\bf E}^{(\rm ge)}=\left(0, \mp \gamma
\frac{GM_om_o}{r^2}, 0\right)\,.
\end{equation}
Hence,
\begin{equation}\label{F3}
f^{(\rm gm)}_y=\left[f_y-f^{(\rm ge)}_y\right]=\pm
v^2\gamma\frac{GM_om_o}{r^2}=\pm vu_x\gamma\frac{GM_om_o}{r^2}\,,
\end{equation}
and therefore
\begin{equation}\label{F4}
{\bf f}^{(\rm gm)}=v\gamma\frac{GM_om_o}{r^2}\left(\mp u_y, \pm
u_x, 0\right)\,.
\end{equation}
This could have been resulted from a GM field as ${\bf f}^{(\rm
gm)}=m_o{\bf u}\times {\bf B}^{(\rm gm)}$, where
\begin{equation}\label{F5}
{\bf B}^{(\rm gm)}=\mp v\gamma \frac{GM_o}{r^2}{\bf \hat{k}}
\end{equation}
is the GM field of the MPM case, or equivalently, relating the GM
field to the relative velocity and the GE field by
\begin{equation}\label{F6}
{\bf B}^{(\rm gm)}={\bf v}\times{\bf E}^{(\rm ge)}\,,
\end{equation}
as expected for the electric analog, see, e.g., Ref.~\cite{res68}.
The same result has been obtained for the MIL case used in
Ref.~\cite{bedfordkrumm85}.

In the above considerations, in order to be consistent with the
approach of Ref.~\cite{kolbenstvedt88}, one should assume that the
speed $v$ to be much smaller than the speed of light. This
assumption just emphasizes that the corresponding results are
unobservable. Incidentally, considering the test mass $m_o$ as a
positive mass charge, the direction of ${\bf B}^{(\rm gm)}$ found
are as if the moving mass charges are negative in analogy with the
electric charge case.

Secondly, we follow the approach of Ref.~\cite{kolbenstvedt88} and
consider a point test mass $m_o$ moving perpendicularly toward a
MIL of constant rest mass charge density $\lambda_o$, under the
influence of its gravitational field in its rest frame, $S'$. The
MIL is assumed to move along its line of direction, as the
$x$--axis, with a constant speed $v\ll1$ with respect to a frame
of reference, say $S$, along the positive $x$--direction. Again,
the orthogonal relative motion of two bodies is considered.

The motion of $m_o$ is determined by $\delta\int(-m_o)ds=0$. As in
Ref.~\cite{kolbenstvedt88}, one can use the fact that in a weak
gravitational potential, $\phi$, the line element, in the SSR
approach, i.e. SR plus the aid of the gravitational time dilation,
is given by
$ds^2=\left(1+2\phi'\right)dt'^{\,2}-\left(dx'^{\,2}+dy'^{\,2}+dz'^{\,2}\right)$.
Using the standard Lorentz transformation for very small $v$, one
obtains
\begin{equation}\label{F7}
-m_ods\simeq-m_o\sqrt{(1+2\phi)dt^2-4\phi v
dtdx-(dx^2+dy^2+dz^2)}\,,
\end{equation}
where $\phi\simeq\phi'$ to the first--order approximation, that is
$m_o$ would follow the same geodesic in either frame, $S$ or $S'$.
As ${\cal L}dt=-m_ods$, one gets
\begin{equation}\label{F8}
{\cal L}\simeq-m_o\sqrt{1+2\phi-4\phi vu_x-u^2}\,.
\end{equation}
Concentrating on the non--relativistic case where $u\ll 1$ and
$\phi \sim u^2$, Eq.~(\ref{F8}), neglecting a constant term
$-m_o$, reads
\begin{equation}\label{F9}
{\cal L}\simeq \frac{1}{2}m_ou^2-m_o\phi+2m_o\phi vu_x\,.
\end{equation}
Comparing this result with the EM analog, one can
deduce\footnote{Here, the vector potential does not explicitly
depend on time.}
\begin{equation}\label{F10}
{\bf B}^{(\rm gm)}=\mbox{\boldmath$\nabla$}\times (2 \phi {\bf
v})=2\, {\bf v}\times (-\mbox{\boldmath$\nabla$}\phi)=2\, {\bf
v}\times {\bf E}^{(\rm ge)}\,.
\end{equation}

In Ref.~\cite{kolbenstvedt88}, the case MPM has been applied,
where the used weak field potential is obviously asymptotically
free, as required. In our applied case, an appropriate key point
is to maintain the asymptotically free assumption radially for the
gravitational field of a long line. This task has been
accomplished in the Appendix. Therefore, using weak field
potential for the MIL case, derived in the Appendix,
Eq.~(\ref{F31}), one gets
\begin{equation}\label{F11}
{\bf B}^{(\rm gm)}=-2v\gamma\,\frac{2G\lambda_o}{y}\hat{{\bf
k}}\,.
\end{equation}
That is, despite a factor of two, Eqs.~(\ref{F10}) and~(\ref{F11})
are equivalent with Eqs.~(\ref{F6}) and~(\ref{F5}), respectively.

In conclusion, comparing Eqs.~(\ref{F6}) and~(\ref{F10}) with the
corresponding results of Refs.~\cite{kolbenstvedt88}
and~\cite{bedfordkrumm85}, respectively, for the same applied
cases, it shows that the strength of the GM field obtained in the
SSR approach is {\bf twice} the SR approach, as we expected.
However, one may still expect to get the correct qualitative
behavior and numerical factor when the full LGR is applied. We
examine this approach in the next section.

\section{Linearized General Relativity and Discussions}
\indent

In this section, we consider the full LGR of the weak field limit
in the following two parts.

Firstly, by using the length contraction as well as the
gravitational time dilation, we employ\footnote{See, for example,
Ref.~\cite{ohanian2nde}.}\
 the metric
$ds^2=(1+2\phi')dt'^{\,2}-(1-2\phi')(dx'^{\,2}+dy'^{\,2}+dz'^{\,2})$
in the rest frame. The static nature of this metric, same as the
Schwarzschild metric, implies that it can only produce GE fields.
However, the existence of cross terms with $dt$ can show that a
stationary space also has a GM field, as in the case of NUT
metric~\cite{ntu1963}, as the generalized Schwarzschild metric, it
has been shown~\cite{dn1966} with the existence of the cross term
$d\varphi dt$.\footnote{For a discussion on NUT space and GM
monopoles see, e.g., Ref.~\cite{ln1998}.}\
 Therefore, in order to extract the GM field from the Lagrangian of
the system for small quantities analogous to the EM case, similar
to the SSR approach, for general cases, including both examples of
the MIL and the MPM, the metric, where the source is moving with a
constant speed $v\ll1$ along the positive $x$--direction, can be
written as
\begin{equation}\label{F12}
ds^2\simeq(1+2\phi)dt^2-(1-2\phi)(dx^2+dy^2+dz^2)-8\phi vdtdx\,.
\end{equation}
With a similar calculation, one gets
\begin{equation}\label{F13}
{\cal L}\simeq (\frac{1}{2}-\phi)m_ou^2-m_o\phi+4m_o\phi vu_x\,.
\end{equation}
Ignoring the second term, $-m_o\phi u^2$, which is a fourth--order
term in $u$, one obtains
\begin{equation}\label{F14}
{\bf B}^{(\rm gm)}=4\, {\bf v}\times {\bf E}^{(\rm ge)}\,,
\end{equation}
see, e.g., Refs.~\cite{rindler79,harris91}.

Interpreting this result in comparison with the light deflection
by a mass where the SR explains only one--half of the deflection,
it has been concluded~\cite{kolbenstvedt88} that the effect of
space curvature to the GM field is equal to the effect of the
gravitational time dilation. On the other hand, the analogy of the
Lagrangian~(\ref{F13}) with the corresponding classical EM case,
which is not already affected by space curvature, is questionable
and doubtful. Indeed, instead of the classical EM case, one should
apply the Einstein--Maxwell equations which, since the EM
energy--momentum tensor, as it is well known, is trace--free, does
exert a sort of constraint on geometry/space
curvature~\cite{farhoudimalek}. The above idea has almost been
studied~\cite{harris91} by taking into account the effect of the
EM field tensor on the geodesic of a particle of mass $m_o$ and
charge $e$ up to the first--order in $v/c$. However, the EM effect
has not practically been considered into the resulting deduction,
instead the analogy with the electromagnetism has just been used.

Implicitly, the existence of the term $-m_o\phi u^2$ in
Eq.~(\ref{F13}) fades a complete analogy between the
Lagrangian~(\ref{F13}) and the EM case. Besides, it has been
claimed~\cite{robertson2007} that Eq.~(\ref{F14}) fails to
correctly describe the free--fall problem in general relativity,
for it apparently does not produce all terms of $v^2/c^2$ order.
However, in this work, we have taken all of these terms into
account and we speculate that the correct result out of the full
linearized equation should be as Eq.~(\ref{F10}). Though, the
equality of the effect of space curvature and the gravitational
time dilation looks to be legitimate.

In the following method, we emphasis on this noticeable point in
order to indicate the best definition amongst those considered in
the literature. Incidentally, there has recently been some debates
in the literature about which one of these definitions leads to
physically correct predictions. Indeed, it has been
claimed~\cite{murphy-etal2007} that Eq.~(\ref{F14}) has been
verified as a source of perturbing acceleration of the lunar
orbit. But, it has been pointed out~\cite{ciufolini2007} that this
effect depends merely on a coordinate frame. Though, it has also
been asserted~\cite{kopeikin2007} that these type of effects are
unobservable, hence the confirmation of Eq.~(\ref{F14}) is
unwarranted~\cite{robertson2007}.

Secondly, we apply the weak field approximation,
$g_{ab}=\eta_{ab}+h_{ab}$ with $|h_{ab}|\ll 1$, and the definition
$\bar{h}_{ab}=h_{ab}-\frac{1}{2}\eta_{ab}h$ with the gauge
condition ${\bar{h}^{ab}{}}_{,b}=0$, as derived in several
textbooks, in order to get the full LGR as
\begin{equation}\label{F15}
\Square\,\bar{h}_{ab}=16\pi GT_{ab}\,,
\end{equation}
where $\Square\equiv\eta^{ab}\partial_a\partial_b$. This equation
is the analogous of the EM potential equation, that is $\Square\,
A^a=4\pi j^a$.

We assume $|\bar{h}_{00}|\gg |\bar{h}_{\alpha\beta}|$ and
$|\bar{h}_{0\alpha}|\gg |\bar{h}_{\alpha\beta}|$, which have been
extensively used in the literature, hence neglecting
$\bar{h}_{\alpha\beta}$, and defining {\setlength\arraycolsep{2pt}
\begin{eqnarray}\label{F16}
&&\bar{h}_{00}:=k_1\phi\,,\cr
&&\bar{h}_{0\alpha}:=k_2A_\alpha\,,
\end{eqnarray} }
where $k_1$ and $k_2$ are constants, therefore the gauge condition
reads
\begin{equation}\label{F17}
k_1\frac{\partial\phi}{\partial t}+k_2\partial_\alpha
A^\alpha=0\,.
\end{equation}
Also, if the gravitoelectromagnetic fields are defined as
\begin{equation}\label{F18}
 {\bf E}^{(\rm ge)}=-\mbox{\boldmath$\nabla$}\phi-\frac{k_2}{k_1}\frac{\partial {\bf
 A}}{\partial t} \hspace{1cm}
 {\rm and}
 \hspace{1cm}{\bf B}^{(\rm gm)}=\mbox{\boldmath$\nabla$}\times {\bf A}\,,
\end{equation}
the gravitoelectromagnetic equations will be
{\setlength\arraycolsep{2pt}
\begin{eqnarray}\label{F19}
 &&\mbox{\boldmath$\nabla$}\cdot {\bf E}^{(\rm
   ge)}=\frac{4}{k_1}\left(4\pi G\rho\right)\,,\cr
 &&\mbox{\boldmath$\nabla$}\cdot\left(\frac{k_2}{k_1}{\bf B}^{(\rm
   gm)}\right)=0\,,\cr
 &&\mbox{\boldmath$\nabla$}\times {\bf E}^{(\rm
   ge)}=-\frac{\partial}{\partial t}
   \left(\frac{k_2}{k_1}{\bf B}^{(\rm gm)}\right)\,,\cr
 &&\mbox{\boldmath$\nabla$}\times
   \left(\frac{k_2}{k_1}{\bf B}^{(\rm gm)}\right)=\frac{\partial {\bf
   E}^{(\rm ge)}}{\partial t}+\frac{4 }{k_1}\left(4\pi G\,{\bf j}\right)\,.
\end{eqnarray} }

An exact analogy with the corresponding EM case, that is the
Lorentz gauge and the Maxwell equations, leads to $k_1=k_2=4$.
This choice is self--consistent, as it regards the special
solution of Eq.~(\ref{F15}), that is
\begin{equation}\label{F20}
\bar{h}_{ab}=4G\int \frac{T_{ab}\left(t-|{\bf x}-{\bf x'}|, {\bf
x'}\right)}{|{\bf x}-{\bf x'}|}d^3x'\,,
\end{equation}
where $T^{00}=\rho$\ and $T^{0\alpha}=j^\alpha$, namely the matter
and current density, as expected.

The weak field approximation, as Eq.~(\ref{F12}), but in a general
case, yields $\bar{h}_{0\alpha}=h_{0\alpha}=-4\phi v^\alpha=4\phi
v_\alpha$, hence, one obtains
\begin{equation}\label{F21}
{\bf A}=\phi {\bf v}\,.
\end{equation}
However, as mentioned before, in order to get
\begin{equation}\label{F22}
{\bf B}^{(\rm gm)}=2\, {\bf v}\times {\bf E}^{(\rm ge)}\,,
\end{equation}
one must replace $\phi$, as the effective gravitopotential, by
$2\,\phi$ arising from the EM analogy. This exertion, as discussed
before, can again be better justified by the analogy that has been
taken between the full linearized equation, which includes the
space curvature, with the classical EM situation. That is, one
should account for the lack of the effect of space curvature in
the EM case~\cite{farhoudimalek}. Similar replacement has also
been performed in the literature, see
Refs.~\cite{rugtar02,mash2003,lim02,mashgl99}, based on the fact
that the linear approximation of general relativity involves a
spin--2 field whereas the electrodynamics involves a spin--1
field. Thus, they have taken the GM charge twice that of the GE
charge.

Implicitly, Eq.~(\ref{F12}) gives $\bar{h}_{00}=4\phi$ and
$\bar{h}_{\alpha\beta}=0$, as we have assumed. Also, the gauge
condition~(\ref{F17}), in the case of Eq.~(\ref{F21}) and when
$\phi$ does not explicitly depend on time, leads to ${\bf
v}\cdot{\bf E}^{(\rm ge)}=0$, which is true in the orthogonal
relative motion cases that we have employed in this work.

On the other hand, if one sets $k_1=4$ and $k_2=2$, as used in the
literature~\cite{rugtar02,mash2003,mashgl99,mash2000}, the same
result of Eq.~(\ref{F22}) will be obtained\rlap.\footnote{Note
that, the appeared sign differences are due to sign conventions.}\
 However, with this choice, the compensation is
that the appearances of the analog EM equations are altered by a
factor of one--half. Though, these factors have also been
justified by the interpretation of the effective GM charge,
however, the point we have raised after Eq.~(\ref{F14}) will not
be clarified in this choice.

The choice $k_1=4$ and $k_2=1$, actually used in
Refs.~\cite{rindler79,harris91}, gives Eq.~(\ref{F14}). Despite
the interpretation made by Ref.~\cite{kolbenstvedt88}, we doubt
this choice to be a correct analog, as mentioned before. Besides,
one can also rewrite the work of Ref.~\cite{harris91} similar to
our work with constants $k_1$ and $k_2$, and, e.g., concludes
$k_1=k_2=4$.

These different values of $k_1$ and $k_2$ are the best and mostly
used ones in the literature. It should be interesting to see
whether these different values give any new information on the
gravitational interactions. In other words, it is instructive if
one makes clear what happens to the equation of motion of a test
particle in the different choices one takes into account. For this
purpose, in the next section, we investigate any effects on the
geodesic equation due to the interplay between the constants $k_1$
and $k_2$ for the GM field.

\section{Geodesic Equation Including Second--Order Approximation}
\indent

In order to examine the GM effects and consequences of these
different definitions on the geodesic equation, we consider the
conditions and the metric (\ref{F12}) with the definitions
(\ref{F16}) and Eq.~(\ref{F21}), for simplicity, in the
$x$--direction. Hence, the weak field potential is
\begin{equation}\label{Geq1}
(h_{ab})=\phi\left(%
\begin{array}{cccc}
  \frac{1}{2}k_1 & -k_2v & 0 & 0 \\
   -k_2v & \frac{1}{2}k_1 & 0 & 0 \\
   0 & 0 & \frac{1}{2}k_1 &  0 \\
   0 & 0 & 0 & \frac{1}{2}k_1 \\
\end{array}
\right)\,.
\end{equation}
As we know, the effects of the GM field on the geodesic equation
can be shown when one maintains the terms of order $\phi^2$,
hence, we will write the relations up to the third--order
approximation. That is, corresponding to the linear approximation
$g_{ab}=\eta_{ab}+h_{ab}$, one gets
\begin{equation}\label{Geq2}
g^{ab}=\eta^{ab}-\eta^{ac}\eta^{bd}h_{cd}+\eta^{ac}\eta^{bd}\eta^{ef}h_{ce}h_{df}+{\cal
O}\bigl(\varepsilon^3\bigr)\,,
\end{equation}
where $\varepsilon\sim\phi\sim v^2$ and, in the rest of this
section, the indices of $h_{ab}$ are not raised by the Minkowski
metric, $\eta^{ab}$, as they usually do up to the first order
approximation.

As before, for a test particle with $u'_x=0=u'_z$ and
$u'^{\,2}_y\sim\varepsilon$, which gives $u_x=v$, $u_y\simeq u'_y$
and $u_z=0$, the geodesic equation,
$\frac{d^2x^a}{d\tau^2}+\Gamma^a{}_{bc}\frac{dx^b}{d\tau}\frac{dx^c}{d\tau}=0$
where $\tau$ is the proper time, in the $y$--direction, yields
 {\setlength\arraycolsep{2pt}
\begin{eqnarray}\label{Geq3}
\frac{d^2y}{dt^2}=-\Bigl[\!\!\!
 &&\Gamma^2{}_{00}+2\left(v\Gamma^2{}_{01}+u_y\Gamma^2{}_{02}\right)\cr
 &&+\left(v^2\Gamma^2{}_{11}+2vu_y\Gamma^2{}_{12}+u_y^2\Gamma^2{}_{22}\right)
  \Bigr]+{\cal O}\bigl(\varepsilon^3\bigr)\,,
\end{eqnarray} }
where we have neglected the term containing $du'_y/dt'$, as a
negligible acceleration using the slow--motion approximation
assumption.

Using Eqs.~(\ref{Geq1}) and (\ref{Geq2}), we get the required
Christoffel symbols, up to the necessary orders, as
{\setlength\arraycolsep{2pt}
\begin{eqnarray}\label{Geq4}
 &&\Gamma^2{}_{00}=\frac{k_1}{4}\frac{\partial\phi}{\partial
  y}+\frac{k_1^2}{8}\phi\frac{\partial\phi}{\partial y}+{\cal
  O}\bigl(\varepsilon^3\bigr)\,,\cr
 &&\Gamma^2{}_{01}=-\frac{k_2}{2}v\frac{\partial\phi}{\partial
  y}-\frac{k_1k_2}{4}v\phi\frac{\partial\phi}{\partial y}+{\cal
  O}\bigl(\varepsilon^3\bigr)\,,\cr
 &&\Gamma^2{}_{11}=\frac{k_1}{4}\frac{\partial\phi}{\partial
  y}+{\cal O}\bigl(\varepsilon^2\bigr)\,,\cr
 &&\Gamma^2{}_{12}=-\frac{k_1}{4}\frac{\partial\phi}{\partial
  x}+{\cal O}\bigl(\varepsilon^2\bigr)\,,\cr
 &&\Gamma^2{}_{22}=-\frac{k_1}{4}\frac{\partial\phi}{\partial
  y}+{\cal O}\bigl(\varepsilon^2\bigr)
\end{eqnarray} }
and null result for $\Gamma^2{}_{02}$ up to the ${\cal
O}\bigl(\varepsilon^3\bigr)$. Hence, Eq.~(\ref{Geq3}) reads
 {\setlength\arraycolsep{2pt}
\begin{eqnarray}\label{Geq5}
\frac{d^2y}{dt^2}=\Bigl(-\frac{k_1}{4}\frac{\partial\phi}{\partial
  y}\Bigr)+\Biggl\{\!\!\!\!
&&-\frac{k_1^2}{8}\phi\frac{\partial\phi}{\partial
 y}+\biggl[\!\Bigl(k_2-\frac{k_1}{4}\Bigr)v^2\cr
&&+\frac{k_1}{4}u^2_y\biggr]\frac{\partial\phi}{\partial
 y}+\frac{k_1}{2}vu_y\frac{\partial\phi}{\partial
 x}\Biggr\}+{\cal O}\bigl(\varepsilon^3\bigr)\,.
\end{eqnarray} }

With the choice of $k_1=4$, the first part on the right hand side,
that is the first order approximation, just reveals the Newtonian
theory, thus, Eq.~(\ref{Geq5}), with $v=u_x$, reads
\begin{equation}\label{Geq6}
 \frac{d^2y}{dt^2}=\Bigl(-\frac{\partial\phi}{\partial
 y}\Bigr)+\Biggl\{-2\phi\frac{\partial\phi}{\partial
 y}+\biggl[\!\Bigl(k_2-1\Bigr)u^2_x
 +u^2_y\biggr]\frac{\partial\phi}{\partial
 y}+2u_x u_y\frac{\partial\phi}{\partial
 x}\Biggr\}+{\cal O}\bigl(\varepsilon^3\bigr)\, ,
\end{equation}
where the second part gives the terms of second--order
approximation and its first term can be justified as the
second--order correction to the Newtonian theory. Its second and
third terms are the GM effect on the same and the transverse
directions, though, in our situation $\phi$ does not depend on
$x$. It also shows that the resulting GM acceleration of a test
particle, i.e. the second term, is proportional to the square of
speed, as has been discussed in Ref.~\cite{robertson2007}. The
proportional coefficients of this term for different choices of
$k_2$ discussed in the previous section, namely $4$, $2$ and $1$,
are $3u^2_x+u^2_y$, $u^2_x+u^2_y=u^2$ and $u^2_y$, respectively.
It will be up to, and may lead easier to, experimental detections
to reveal which one of these values, and hence definitions, may
lead to physically correct predictions. However, the existence of
these coefficients in geodesic equations may also be employed in
tuning the numerical results of experiments.

\section{Conclusion}
\indent

We have investigated apparent GM fields, which can be removed with
suitable choice of coordinates, or which can be arisen on the
basis of the intriguing interplay between geometry and dynamics.
We have demonstrated that the strength of the GM field obtained in
the SSR approach is twice the SR approach, and we have argued that
the results of LGR equations must be the same as SSR approach. Our
argument is mainly based on the fact that the corresponding EM
case is not affected by space curvature, which is shown to have
the same effect on the GM field as the gravitational time
dilation. By this, we have justified that one must replace the
effective gravitopotential, $\phi $, by $2\phi $, and thus achieve
an exact analogy with the corresponding EM equations for the GM
fields. Hence, the best definition for the gravitoelectromagnetic
equations with coefficients $k_1$ and $k_2$ are obtained when the
values for both of them are set equal to four. Also, our
stimulated hope in deriving the geodesic equation including
second--order approximation is that more theoretical understanding
of this procedure may lead to a practical detection of this
phenomenon.

\section*{Appendix}
\indent

In this Appendix, we derive an asymptotically flat metric, in the
radial direction, for the MIL case, although, a sort of general
treatments can be found in Ref.~\cite{stephani}.

A most general cylindrically symmetric static metric in four
dimensions with signature $-2$ can be written in the canonical
form, in a given frame as\footnote{The $z$ direction in this
Appendix corresponds to the $x$ direction in the text.}
\begin{equation}\label{F23}
ds^2=e^{2F(\rho)}dt^2-e^{2H(\rho)}d\rho^2-\rho^2d\varphi^2-e^{2U(\rho)}dz^2\,,
\end{equation}
where the existence of a function $U(\rho)$ and the factor of two
in the exponentials are for later convenience.

The unknown functions $F(\rho)$, $H(\rho)$ and $U(\rho)$ can be
determined using the Einstein vacuum equations. After some
calculations, one gets the following differential equations
{\setlength\arraycolsep{2pt}
\begin{eqnarray}\label{F24}
&&F''(\rho)+\frac{F'(\rho)}{\rho}=0\,,\cr
&&U''(\rho)+\frac{U'(\rho)}{\rho}=0\,,\cr
&&H'(\rho)=F'(\rho)+U'(\rho)
\end{eqnarray} }
and
\begin{equation}\label{F25}
F''(\rho)+U''(\rho)-F'(\rho)U'(\rho)-\frac{F'(\rho)+U'(\rho)}{\rho}=0\,,
\end{equation}
where the prime represents derivative with respect to $\rho$.
Eqs.~(\ref{F24}) can easily be used to get
{\setlength\arraycolsep{2pt}
\begin{eqnarray}\label{F26}
&&F(\rho)=F_o\ln\frac{\rho}{a}\,,\cr
&&U(\rho)=U_o\ln\frac{\rho}{b}\,,\cr
&&H(\rho)=H_o+F(\rho)+U(\rho)\,,
\end{eqnarray} }
where $F_o$, $U_o$, $H_o$, $a$ and $b$ are constants of
integration, Eq.~(\ref{F25}) implies
\begin{equation}\label{F27}
2(F_o+U_o)+F_oU_o=0\,.
\end{equation}
This is actually due to the contracted Bianchi identities that
makes Eqs.~(\ref{F24}) and~(\ref{F25}) not to be independent.

For weak gravitational fields, one can assume $F_o$ and $U_o$ to
be small. Hence, neglecting the term $F_oU_o$ in Eq.~(\ref{F27}),
one gets $F_o\approx -U_o$. In this approximation, the line
element~(\ref{F23}) reads
\begin{equation}\label{F28}
ds^2\simeq
\Bigl(1+2F_o\ln\frac{\rho}{a}\Bigr)dt^2-h_od\rho^2-\rho^2d\varphi^2-\Bigl(1-2F_o\ln\frac{\rho}{b}\Bigr)dz^2\,,
\end{equation}
where  $H(\rho)=H_o+F_o\ln (b/a)\equiv h_o$ is actually a
constant.

In a local frame of reference, using the standard Lorentz
transformation between inertial frames, that is $x\rightarrow x$,
$y\rightarrow y$, $z\rightarrow \gamma(z-vt)$ and $t\rightarrow
\gamma(t-vz)$, one gets a stationary metric, namely
{\setlength\arraycolsep{2pt}
\begin{eqnarray}\label{F29}
&ds^2\simeq
&\Bigl(1+2\gamma^2F_o\ln\frac{\rho}{a}\Bigr)dt^2-h_od\rho^2-\rho^2d\varphi^2\cr
&&-\Bigl(1-2\gamma^2F_o\ln\frac{\rho}{b}\Bigr)dz^2+4\gamma^2vF_o\ln\frac{a
b}{\rho^2}dz dt\,,
\end{eqnarray} }
where we have neglected the third--order terms containing
$v^2F_o$.

Now, let us use the above results for the MIL case. First of all,
our \emph{priori} assumption that the space--time would be static
can be justified by the constant linear velocity of the MIL.
Besides, it is also clear that a static space--time can be evident
only in its adapted coordinate system, and not in a general
coordinate, e.g., Eq.~(\ref{F29}). However, one should note that,
as has been discussed in Ref.~\cite{bonnor2006} and references
therein, a rotating line mass and/or mixed time--independent
electric and magnetic fields, in general, cause rotational effects
in space--time, and hence, the space--time will be stationary but
not static.

In order for the metric to be asymptotically flat in the radial
direction, we assume the following simple case of $b=a$ and
$h_o=1$, hence the metric~(\ref{F29}) reads
{\setlength\arraycolsep{2pt}
\begin{eqnarray}\label{F30}
&ds^2\simeq &
\Bigl(1+2\gamma^2F_o\ln\frac{\rho}{a}\Bigr)dt^2-d\rho^2-\rho^2d\varphi^2\cr
&&-\Bigl(1-2\gamma^2F_o\ln\frac{\rho}{a}\Bigr)dz^2-8\gamma^2vF_o\ln\frac{\rho}{a}dz
dt\,.
\end{eqnarray} }
The metric ~(\ref{F30}), for the radial distance $\rho=a$,
obviously gives the Minkowski flat metric. Henceforth, to comply
with the required assumption, we assume that this should be true
when $\rho>a$ as well, i.e. a cutoff has been applied. That is,
$a$ must be a sufficiently large perpendicular distance from the
line where, on and beyond it, the gravitational field tends to
zero\rlap,\footnote{Evidently, $a$ depends on the line mass charge
density.}\
 and actually, the
metric~(\ref{F30}) is valid for $\rho\leq a$.

To determine $F_o$, one should note that the gravitational
potential for a long line mass charge density at rest,
$\lambda_o$, is $\phi'=\phi'_o+ 2G\lambda_o \ln \rho$. Choose,
$\phi'_o=-2G\lambda_o \ln a$, and amend equations when the line is
moving, that is replace $\lambda_o$ by $\gamma\lambda_o$. For the
weak field case, where $g_{00}=1+2\phi$, one obtains\footnote{If
one keeps the third order terms containing $v^2F_o$, one obtains
$F_o=2G\lambda_o/\gamma(1+v^2)$.}\
 $F_o=2G\lambda_o/\gamma$. That is, the asymptotically flat
metric, in the radial direction for the MIL case in the frame $S$,
when $\rho\leq a$, is given by
 {\setlength\arraycolsep{2pt}
\begin{eqnarray}\label{F31}
&ds^2\simeq & \left(1+4\gamma
  G\lambda_o\ln\frac{\rho}{a}\right)dt^2-d\rho^2-\rho^2d\varphi^2\cr
&&-\left(1-4\gamma G\lambda_o\ln\frac{\rho}{a}\right)dz^2-16\gamma
  G\lambda_o v\ln\frac{\rho}{a}\, dz dt\,.
\end{eqnarray} }

\end{document}